\documentstyle[aps,prl,floats]{revtex}
\begin{document}
\input psfig.sty
\draft
\twocolumn[\hsize\textwidth\columnwidth\hsize\csname@twocolumnfalse%
\endcsname

\title{\bf Extraction of Electrostrong Parameters
of $N^{*}(1520)$ from Eta Photoproduction}

\author{Nimai C. Mukhopadhyay and Nilmani Mathur}

\address
{Department of Physics, Applied Physics and Astronomy \\
Rensselaer Polytechnic Institute \\
Troy, New York 12180-3590}

\date{\today}

\maketitle
\vskip 12pt
%\vspace{-0.2in}

\begin{abstract}
Recently obtained polarized target and photon asymmetry data in
eta photoproduction are shown to be very powerful, in conjunction with
the differential cross-section data, in yielding model-insensitive
constraints on the electrostrong parameters for the excitation and decay of
the $N^{*}(1520)$ resonance. The extracted ratio of its
electromagnetic helicity amplitudes,
$A_{3\over 2}/A_{1\over 2}$, provides a critical test for the QCD-inspired
hadron models.
\end{abstract}

$\quad\quad\quad\quad\quad\,\,${\it Keywords}: $\eta$ meson, $N^{*}(1520)$
and $N^{*}(1535)$ resonances,
Polarization observables, Effective Lagrangian
]

\narrowtext

Studies of the electromagnetic \cite{1} and weak \cite{2} transition
amplitudes to various resonance states of the nucleon ($N$) as a function of
the square of the four-momentum transfer, $q^{2}$, is a powerful way to
explore the chromodynamic structure of the nucleon. The real photon
point, for which $q^{2} = 0$, is one end
of the domain of
non-perturbative QCD, which continues until some large $q^{2}$, as yet
unknown, which marks
the onset of perturbative QCD (pQCD). For the $N$ to
$N^{*}$ transitions, with
the spin of the $N^{*}$ being ${3\over 2}$, there are two helicity
amplitudes,
$A_{1\over 2}$ and $A_{3\over 2}$, for real photon excitations. In contrast
to the pQCD domain, where counting rules yield
$A_{1\over 2} >>  A_{3\over 2}$ \cite{3}, the
non-perturbative region is characterized by a large helicity
violation \cite{1}.
In this Letter, we shall study this
in the $N$ to $N^{*}(1520)$ real photon transition, via the reaction
\begin{equation}
\gamma + p \quad \longrightarrow \quad p + \eta \, ,
\end{equation}
with the photon lab energy from the eta photoproduction threshold of
707 MeV, up to about 900 MeV, dictated by the availability of data and
relative
dynamical simplicity.

Though a well-established resonance from the analyses of
the pion-nucleon scattering and pion photoproduction,
the electromagnetic properties of $N^{*}(1520)$, are yet to be studied
experimentally via complementary
reactions such as (1) and fully understood in the framework of
the QCD-inspired models. Our interest in the reaction (1) is also enhanced
by the recent availability of high quality data from different photon
factories: differential cross-section data from the Mainz microtron
\cite{4},
the polarized target asymmetry (PTA) from the upgraded electron facility at
Bonn \cite{5} and a precise data
set, just released, on the  polarized photon asymmetry (PPA)
from the French laser light source, GRAAL \cite{6}. We show below that
a combination of these observables provide a {\it powerful constraint} on
relatively small effects from the excitation of the $N^{*}(1520)$
resonance and its decay, {\it amplified  by}  the interference with the
dominant contribution of the $N^*(1535)$ resonance. There are also subtle
issues arising from the nodal structures \cite{7} of
these observables. Finally,
we discuss implications of these data
on the non-perturbative QCD violation of the helicity conservation in
the electromagnetic process $N \rightarrow N^{*}(1520) $, and in particular,
on the testing of various QCD-inspired models \cite{8} of hadrons.

Absent lattice gauge theoretic estimates, our theoretical knowledge of
helicity amplitudes for the baryon resonance excitation comes from the
QCD-inspired models \cite{8}. Current level of their uncertainties for the
helicity amplitudes for the $N \rightarrow  N^{*}(1520)$  excitation is as
follows : the $A_{1 \over 2}$ amplitude for the proton target ranges
from $-13$ to $-51$, in units of $10^{-3}$GeV$^{1/2}$, in various versions
\cite{8} of the constituent quark model, while the $A_{3 \over 2}$
amplitude is predicted to be in the range 117 to 173 in the same units.
The 1996 PDG \cite{9} values of these amplitudes from pion photoproduction
quote relatively small errors, but they do not include theoretical
uncertainties of the position, branching ratios and width of the resonance.
The relatively large uncertainties in the partial width $\Gamma_{\eta}$
($\sim 0.14$ MeV) of
the resonance $N^{*}(1520)$ to decay into $\eta N $ channel
and the total width, $\Gamma$, known to be between 110 to 135 MeV, result
in errors in extracting $A_{i}$ from the photoproduction data,
{\it much bigger} than the errors quoted by the PDG.
An attempt
to get these amplitudes from the $(\gamma,\pi\pi)$ reaction has yielded
the $A_{3\over 2}$ amplitude only within a factor of two \cite{10},
with no meaningful constraint on $A_{1\over 2}$.

We shall extract from the process (1) the parameters $\xi_{i} (i = {1 \over
2},
{3 \over2} )$ defined as \cite {11} :
\begin{equation}
\xi_{i} \quad = \quad \sqrt{\chi \Gamma_{\eta}} A_{i}/\Gamma ,
\end{equation}
where $\chi$ is a kinematic parameter, $ Mk/(q M_{R})$, $k$ and $q$ are
the photon and eta meson momenta in the $\eta N\, cm$ frame, $M$ and $M_{R}$
are the nucleon and the resonance masses. One of our findings is
that the new cross-section and polarization data of the reaction (1),
{\it taken together}, give us precise estimates of the quantities
$\xi_{1/2}$ and $\xi_{3/2}$ for $N^{*}(1520)$ for the first time from
the reaction (1). Thus the quantity
$\xi_{3/2}/\xi_{1/2}$ yields an estimate of the ratio of electromagnetic
helicity amplitudes $A_{3/2}/A_{1/2}$ {\it essentially independent
of uncertainties of the strong interaction parameters}, which drop out in
the ratio. This is a {\it crucial result} of this Letter, of substantial
value to distinguish among competing hadron models.

Our theoretical tool for analysis of the reaction (1) is the effective
Lagrangian approach,
which consists, in the tree approximation
of the $s$- and $u$- channel nucleon and resonance
Born terms and the $t$-channel vector meson ($\rho$ and $\omega$) exchanges
\cite{1,11}. Dominant contributions for eta photoproduction around
$W \sim 1.3 $ GeV are well-studied, consisting of the nucleon Born terms
and $s$-channel excitation of $N^{*}(1535)\, {1\over 2}^{-}, T = {1 \over 2}$
resonance \cite {11,12}. Our goal here is to get at the relatively small
contributions from the excitation of $N^{*}(1520),  {3\over 2}^{-},
T = {1 \over 2}$ resonance. We cannot do that from the differential
cross-section data alone, even though there is a hint \cite{1,4,12,13}
of its presence from these data. It is a  {\it combination}
of these differential cross-section data with
the recently gathered data \cite{5,6} on polarization observables that
allows us to put powerful constraints on $N^{*}(1520)$ amplitudes.
The PPA turns out to be rather insensitive to $A_{1/2}$,
thereby giving us a better fix on $A_{3/2}$,
while the differential cross-section and the PTA help us to constrain the
$A_{1/2}$ amplitude.

We shall now briefly discuss the general structure of the interaction
Lagrangian for the $ {3\over 2}^{-}, T = {1 \over 2}$ resonance excitation.
The strong and electromagnetic pieces are \cite{11}:
\begin{eqnarray}
L_{\eta N R} &=& {g_{R}\over \mu} \bar{R}^{\mu}
\theta_{\mu\nu}(Z) \gamma_{5} N \partial^{\nu} \eta \, + \, h.c.,\\
L^{1}_{\gamma N R} &=& {i e \over 2 M} \bar{R}^{\mu}
\theta_{\mu\nu}(Y) \gamma_{\lambda} (G^{1}_{s} + \tau_{3} G^{1}_{v})
N F^{\lambda\nu}\, + \, h.c.,\\
L^{2}_{\gamma N R}  &=& - { e \over 4 M^{2}} \bar{R}^{\mu}
\theta_{\mu\nu}(X) (G^{2}_{s} + \tau_{3} G^{2}_{v})
 (\partial_{\lambda}N) F^{\nu\lambda} \nonumber\\
&&\hspace{2in} + \, h.c.,
\end{eqnarray}
where the tensor $\theta_{\mu\nu}(A)$ is defined as follows \cite{14}:
\begin{equation}
\theta_{\mu\nu}(A)\quad = \quad g_{\mu\nu} - [{1 \over 2} (1 + 2A)]
\gamma_{\mu} \gamma_{\nu}.
\end{equation}
Parameter $A$ is not {\it a priori}
known and it must be determined from the fits to the data on the reaction
(1).
$R$ is the vector-spinor field for the spin-${3\over 2}$ resonance; the
resonant three-point couplings for the proton target, $g_{R}$,
$G^{i}_{P} = G^{i}_{s} + G^{i}_{v}$ are all to be determined from the fits to
the data of the reaction (1); $ F^{\mu\nu}$ is the electromagnetic field
tensor representing the external real photon field. In the broadest fit
we have attempted, we have {\it nine} effective parameters : in the
non-resonant part  of the amplitude, these are the eta-nucleon coupling
and two vector meson couplings; in the resonant sector we have one helicity
amplitude for  $N^{*}(1535)$ and two helicity amplitudes $A_{1\over 2}$,
$A_{3\over 2}$ for $N^{*}(1520)$ excitation and three ``off-shell''
parameters, $ X, Y ,Z \cite{14}$.
We use the CERN routine MINUIT \cite{15} for these fits. This helps
us to get the global $\chi^2$ minimum in the fitting process.

We start with the expressions for the observables of our interest
in terms of the helicity amplitudes $H_{i} (i = 1,2,3,4)$ and write
them in terms of
multipole amplitudes up to $d$-waves in the $\eta N$ channel
\cite{1}, exhibiting only terms involving the dominant
$E_{0+}$ multipole.
Thus, the differential cross-section $d\sigma \over d\Omega$,
the PPA $[ \Sigma ] $ and the PTA $ [ T ] $ are given by \cite{1,16}
\begin{eqnarray}
{d\sigma\over d\Omega}& = & {|{\vec {q}}|
\over  2\,|{\vec {k}}|}\,\sum_{i = 1}
^{i = 4} |H_{i}|^2\nonumber\\
&&\hspace{-0.4in}=\,{|{\vec{q}}|
\over |{\vec {k}}|}\,{\biggl[} E^{2}_{0+}
- {\hbox{Re}}{\left\{E^{*}_{0+} (E_{2-}
-3M_{2-} + 3M_{2+} + 6E_{2+})\right\}}\nonumber\\
&&\quad\quad+\,\,2 \,{\hbox {cos}} \theta\,{\hbox{Re}}\, {\left\{E^{*}_{0+}
( 3 E_{1+} + M_{1+} - M_{1-})
\right\}}\nonumber\\
&&\hspace{-0.3in}\,+\,\,3\, {\hbox {cos}}^{2} \theta\,
{\hbox{Re}}\,\{E^{*}_{0+} ( E_{2-} - 3M_{2-}
\,+\, 6 E_{2+} + 3M_{2+})\}{\biggr]},\\
&&\hspace{-0.2in}{d\sigma\over d\Omega}\,\Sigma \,\,=\,\,
{|{\vec{q}}| \over  |{\vec{k}}|}\,
{\hbox{Re}}\, {\left\{H_{1} H_{4}^{*} - H_{2} H_{3}^{*}\right\}}\nonumber\\
&&\hspace{-0.3in}=\,-\,{|{\vec{q}}|
\over |{\vec{k}}|}\,3 \,{\hbox {sin}}^{2} {\theta}\, {\hbox{Re}} {\biggl[}
E^{*}_{0+}(M_{2+}-E_{2+}-M_{2-}-E_{2-}) {\biggr]},\\
&&\hspace{-0.2in}{d\sigma\over d\Omega}\,T \,\,=\,\,  {|{\vec{q}}|
\over  |{\vec{k}}|}\,
{\hbox{Im}} \, {\left\{H_{1} H_{2}^{*} + H_{3} H_{4}^{*}\right\}}\nonumber\\
&&\hspace{0.22in}=\,\,{|{\vec{q}}|
\over |{\vec{k}}|}\,3\,{\hbox {sin}} {\theta}\, {\hbox{Im}} {\biggl[}
E^{*}_{0+}(E_{1+} - M_{1+})\nonumber\\
&&\quad\quad
+\, E^{*}_{0+}(4E_{2+}-4M_{2+}-M_{2-}-E_{2-})
\, {\hbox {cos}} \theta {\biggr]}.
\end{eqnarray}
\noindent
%%%
Above we are omitting the interference terms between $p$- and $d$-wave
multipoles for brevity, although they are included in our calculation.
They are crucial to understand
sensitivity to the $A_{1/2}$ helicity amplitude in our chosen
observables. This subtle interference effect of the $p$-wave multipoles is
ignored in a recent analysis \cite{13} of Tiator {\it et al.},
as is the complex Lagrangian structure of the spin-$3\over 2$ vertex in
Eqs.(3)-(6).

In the ``second'' resonance region, around $W \sim 1.3 $ GeV,
of interest here, the dominant
multipole for eta photoproduction  is $E_{0+}$,
and its primary contribution is  from the excitation of
the  $N^{*}(1535)$ resonance \cite{12}. Both the $E_{0+}$ and $M_{1-}$
multipoles also receive contributions from the spin-$1 \over 2$
sector of the $N^{*}(1520)$, often referred to as the off-shell sector
\cite{14} of the spin-$3 \over 2$ resonance. This  is controlled
by the parameters $X, Y$ and $Z$ introduced earlier in Eqs. (3)-(5).
The multipoles $E_{2-}$ and $M_{2-}$, in which $N^{*}(1520)$ is
resonant, are relatively small in the energy region of our interest,
but are retained for an important reason. Their effects are enhanced
by the interference with the large $E_{0+}$ multipole (Eq.7-9),
in contrast to pion photoproduction where no single multipole stands out.

The differential cross-section $d\sigma \over d\Omega$ of the reaction (1),
recently determined at the Mainz Microtron \cite {4}, is very flat near
the eta photoproduction threshold characteristic \cite{1} of the dominance of
the $E_{0+}$ multipole and $N^{*}(1535)$  excitation. As the photon energy
increases, the differential cross-section begins to deviate from near
isotropy and shows angular dependence (Figs.1, first column).
This has been interpreted \cite{11,13} as
a complicated effect of a combination of nucleon Born terms and the role of
the $N^{*}(1520)$ excitation. However, the best fit of the Mainz data
alone {\it misses} the sign of the PTA (Figs.1, second column)
and {\it cannot} reproduce the magnitude of the PPA  \cite{6} (Figs.1,
third column).
It is only a {\it combination of these three data sets},
encompassing broad energy range,
[4-6] that results in the acceptable fits to all these diverse data,
a sample of which is represented by the solid lines.
There is also incompatibility between the low energy PTA data
and other observables.
If we force a fit to the low-energy data, we cannot use that fit (dashed
lines) to describe the higher energy data sets. We should recall that
isobar model fits of the low-energy PTA data have been so far unsuccessful
\cite{13}.

This brings us to the subject of nodal structure of the observables as
fingerprints of resonance contributions \cite{7}. The nodal structure
anticipated by Saghai and Tabakin \cite{7} for {\it pure} s- and d-wave
resonances, follows from a
$sin\theta \cdot cos\theta$ distribution for the PTA.
However, the $s$-wave interfering with the $p$ and $d$-wave
multipoles, predicted in our effective Lagrangian approach, spoils this
simple expectation. The data at higher energies support the latter, with no
node appearing in the PTA (Figs.1, second column).
Such deviations are indicative of subtle roles of background
contributions.

We have to deal with the problem of broad range of parameters for
resonances in the PDG compilation. We adopt the following strategy
to do our fits.
For a particular set of $N^{*}(1520)$ parameters, we vary properties
of $N^{*}(1535)$, such as mass, width  etc., within the permitted
PDG 1996 boundaries. We then
change the parameter set of $N^{*}(1520)$ and repeat the procedure.
In this way we cover many possible parameter sets of $N^{*}(1520)$
and $N^{*}(1535)$. Shown in Table I are 
the parameter sets for which we got the 
$\chi^{2}$ per degree of freedom around 1.3, our lowest $\chi^2$ level.
The fitted values of the parameters for the $N^{*}(1520)$ resonance correlate
strongly with the properties of $N^{*}(1535)$, such as its position, total
width and the eta-nucleon partial width, as shown in Table I.
However, {\it the electrostrong parameters for 
the  $N^{*}(1520)$, $\xi^{D}_{3/2}$
and $\xi^{D}_{1/2}$ are relatively stable}.
By taking a broad band of allowed parameters around the lowest value of the
$\chi^{2}$ per degree of freedom, $(\sim 1.30)$, we can ascertain
the following set of $N^{*}(1520)$
parameters :
\begin{eqnarray}
\xi^{D}_{3\over 2} &=& 0.165\pm 0.015 \pm 0.035 ,\\
\xi^{D}_{1\over2} &=& -0.065\pm 0.010 \pm 0.015 .
\end{eqnarray}
where the first error is statistical and the second one is systematic.
The systematic error reported here is due to uncertainties in positions,
branching ratios and decay widths of the resonances (Table I), while the
statistical error is obtained from the fitting program MINUIT \cite{15}.

In the ratio of these two parameters, strong interaction uncertainties drop
and we can determine, in a nearly {\it model independent} manner, a ratio of
the helicity amplitudes $A_{3/2}$ to $ A_{1/2}$  for $N^{*}(1520)$ :
\begin{equation}
A_{3/2}/A_{1/2} = - 2.5 \pm 0.2 \pm 0.4.
\end{equation}
This ratio has been reported by the 1996 PDG as $-6.9 \pm 2.6$ \cite{8},
from pionic processes.{\it Ours is the first determination
of this quantity from eta photoproduction.} The importance
of the difference between our value and the PDG one will become apparent
below, when we compare it with model estimates. This ratio could
only be determined here by exploiting both the differential cross-section
\cite{4} and polarization observables \cite{5,6} together.

\begin{figure}[h]
\vskip 3in
\includegraphics{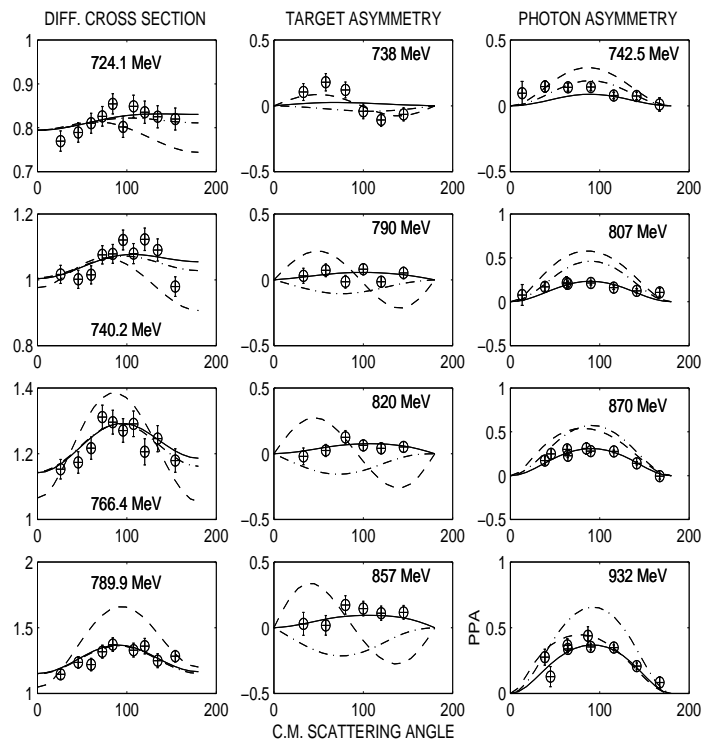}
\vskip 1.0in
\caption{Differential cross-section and polarization observables
in our effective Lagrangian fit with different choices of energy segments.
Solid : Fitting Mainz [4], Bonn [5] and GRAAL [6] data
upto 790, 895 and 932 MeV respectively; dash-dot: fitting Mainz data
alone; dash: fitting a low-energy truncation of the
Mainz and Bonn data, upto 750 MeV.}
\end{figure}

We now discuss the significance of the ratio (12) from the point of
view of the structure of the nucleon and $N^{*}(1520)$. It should
be zero in the pQCD regime \cite{3}. We are
clearly dealing here with non-perturbative physics that is altering the
helicity structure dramatically in the $q^{2} \rightarrow 0$ limit.
This helicity structure is very model-dependent.
Many topical models \cite{8} of baryon structure in the
literature attempt to address this. A sample of their
predictions is given in Table II. We note that the non-relativistic
quark model of Isgur and Koniuk \cite{8} yields a value $-5.6$, while 
Li and Close \cite{8}, taking into account effects of color hyperfine
interaction in the quark transition operators,
are estimating its value in between $-2.5$ to $-4.5$. Our phenomenologically 
extraxcted value is in excellent agreement with the prediction ($-2.5$) of 
Bijker {\it et al.}\cite{8}. Their model deals with a dynamical 
$U(7)$ symmetry of the nucleon structure that has a oblate top spectrum.
For the $N^{*}(1520)$ excitation at $q^{2} \rightarrow 0$,
we have confirmed the helicity inequality
$|A_{1/2}|\, < \, |A_{3/2}|\,, $
from the data, in contrast to the
inequality
$|A_{1/2}|\, >> \, |A_{3/2}|\,$ expected at high $q^{2}$.

\begin{table}[h]
\caption{The electrostrong parameters as determined for $N^{*}(1535)$
and $N^{*}(1520)$ for different sets of resonance position $M_{R}$,
total width $\Gamma$ and $\eta N$ branching ratio.
The sets are : a1 = 1544, 212, 0.45; a2 = 1535, 185, 0.45;
b1 = 1515, 135, 0.0012; b2 = 1530, 135, 0.0012;
b3 = 1530, 110, 0.0012, b4 = 1520, 120, 0.0012.}
\vspace{0.1in}
\begin{tabular}{||c|ccccc||} %\hline
Parameter & $\chi^{2}$ & $\xi^{S}_{1/2}$ & $\xi^{D}_{1/2}$  & $\xi^{D}_{3/2}$
& $R$  \\
Set &per d.f. &&&&
\\
\hline
$a1, b1 $&$1.349$&$2.221$&$-0.053$&$0.145$&$-2.73$\\
&&&$\pm 0.009$&$\pm 0.017$&$\pm 0.56$\\
\hline
%&&&&&\\
$a1, b2 $&$1.308$&$2.218$&$-0.075$&$0.186$&$-2.48$\\
&&&$\pm 0.011$&$\pm 0.021$&$\pm 0.46$ \\
\hline
%&&&&&\\
$a1, b3 $&$1.310$&$2.219$&$-0.080$&$0.198$&$-2.47$\\
&&&$\pm0.011$&$\pm0.021$&$\pm 0.43$\\
\hline
$a1, b4 $&$1.329$&$2.221$&$-0.063$&$0.167$&$-2.65$\\
&&&$\pm0.009$&$\pm0.019$&$\pm 0.48$\\
\hline
%&&&&&\\
$a2, b1$ &$1.329$&$2.308$&$-0.051$&$0.132$&$-2.59$\\
&&&$\pm0.005$&$\pm0.011$&$\pm 0.33$\\
\hline
%&&&&&\\
$a2, b2$ &$1.301$&$2.307$&$-0.074$&$0.172$&$-2.32$\\
&&&$\pm0.007$&$\pm0.015$&$\pm0.30$\\
\hline
%&&&&&\\
$a2, b3$ &$1.295$&$2.307$&$-0.074$&$0.181$&$-2.44$\\
&&&$\pm 0.007$&$\pm 0.015$&$\pm 0.31$\\
\hline
$a2, b4$ &$1.308$&$2.307$&$-0.059$&$0.152$&$-2.58$\\
&&&$\pm 0.006$&$\pm 0.013$&$\pm 0.34$\\
\end{tabular}
\end{table}
\begin{table}[h]
\vspace{-0.1in}
\caption{The ratio $R = \xi^{D}_{3/2}/\xi^{D}_{1/2}$ as predicted in various
models
[8] for the $N^{*}(1520)$ excitation and decay into $\eta N$, and as
determined in a model-independent manner from this work.}
\vspace{0.1in}
\begin{tabular}{|cccccc|} %\hline
Isgur-&Capstick&Li-&Bijker&Inferred from&This\\
Koniuk&&Close&{\it et al.}&PDG 96 &work\\
\hline
$-5.5$6&$-8.93$&$-2.49$&$-2.5$&$-6.9$&$-2.5$\\
&&to$-4.86$&&$\pm2.6$&$\pm 0.2 \pm 0.4$\\
\end{tabular}
\end{table}

In summary, the recent experimental advances in the study of photoproduction
of eta mesons in the second resonance region ($W \sim 1.3$ GeV)  has
immediate
theoretical pay-off for the knowledge of electromagnetic amplitudes that
excite $N^{*}(1520)$. Even though this resonance is a relatively minor
player in this reaction, a combination of the differential cross-section
and polarization data, coming out of the recent experiments at the photon
facilities, have allowed us to infer in a nearly
 model-independent way the value of
the ratio $A_{3/2}/A_{1/2}$, which is predicted to
be negative in the QCD-inspired models, but is strongly model-dependent.
The magnitude of this ratio is selective among these topical models, and is
very different from being zero, expected in the pQCD regime.
New theoretical work is needed to explore this
ratio on the lattice. On the experimental side, electroproduction of
pseudoscalar mesons, at facilities like the CEBAF, will
throw new light on the $q^{2}$ developments of these
helicity amplitudes and their longitudinal partner.

We thank R.M. Davidson,  M. Benmerrouche and J.-F. Zhang for many discussions
and numerical helps. We are grateful to A. Bock, J.-P. Didilez, B. Krusche,
B. Saghai and L. S$\ddot{\hbox{o}}$z$\ddot{\hbox{u}}$er for
communicating to us preliminary
results from eta experiments at Bonn, GRAAL and Mainz. We  also thank
J. A. Gomez-Tejedor, E. Oset, B. Saghai, F. Tabakin and R. Workman
for clarifying
related theoretical issues. Our research is
supported by the U. S. Department of Energy.

%\vspace{-0.24in}

\end{document}